\documentclass[aps,prb,twocolumn,floatfix,amsmath,amssymb,showpacs,
               superscriptaddress]{revtex4-1}
\usepackage{graphicx}
\usepackage{color}
\usepackage{dcolumn}
\usepackage{bm}
\usepackage{epsfig,psfrag,amsmath,amssymb,float}
\input{epsf}
\usepackage{hyperref}
\usepackage[percent]{overpic}
\hypersetup{
    colorlinks=true,
    linkcolor=blue,
    citecolor=blue,
    filecolor=magenta,      
    urlcolor=cyan,
}
\usepackage{array}
\newcolumntype{C}[1]{>{\centering\arraybackslash$}p{#1}<{$}}

\setcitestyle{square}
\newcommand{\pdag}{{\phantom{\dagger}}}
\bibpunct{[}{]}{,}{n}{}{}
\begin{document}
\title{Density matrix renormalization group study of a three-orbital Hubbard model with spin-orbit coupling in one dimension}

\author{Nitin Kaushal}
\affiliation{Department of Physics and Astronomy, The University of 
Tennessee, Knoxville, Tennessee 37996, USA}
\affiliation{Materials Science and Technology Division, Oak Ridge National 
Laboratory, Oak Ridge, Tennessee 37831, USA}

\author{Jacek  Herbrych}
\affiliation{Department of Physics and Astronomy, The University of 
Tennessee, Knoxville, Tennessee 37996, USA}
\affiliation{Materials Science and Technology Division, Oak Ridge National 
Laboratory, Oak Ridge, Tennessee 37831, USA}

\author{Alberto Nocera}
\affiliation{Department of Physics and Astronomy, The University of 
Tennessee, Knoxville, Tennessee 37996, USA}
\affiliation{Materials Science and Technology Division, Oak Ridge National 
Laboratory, Oak Ridge, Tennessee 37831, USA}

\author{Gonzalo Alvarez}
\affiliation{Center for Nanophase Materials Sciences, Oak Ridge National 
Laboratory, Oak Ridge, Tennessee 37831, USA}
\affiliation{Computational Science and Engineering Division, Oak Ridge National 
Laboratory, Oak Ridge, Tennessee 37831, USA}

\author{Adriana Moreo}
\affiliation{Department of Physics and Astronomy, The University of 
Tennessee, Knoxville, Tennessee 37996, USA}
\affiliation{Materials Science and Technology Division, Oak Ridge National 
Laboratory, Oak Ridge, Tennessee 37831, USA}

\author{F. A. Reboredo}
\affiliation{Materials Science and Technology Division, Oak Ridge National 
Laboratory, Oak Ridge, Tennessee 37831, USA}

\author{Elbio Dagotto}
\affiliation{Department of Physics and Astronomy, The University of 
Tennessee, Knoxville, Tennessee 37996, USA}
\affiliation{Materials Science and Technology Division, Oak Ridge National 
Laboratory, Oak Ridge, Tennessee 37831, USA}

\date{\today}

\begin{abstract}
Using the Density Matrix Renormalization Group technique we study the effect of spin-orbit coupling 
on a three-orbital Hubbard model in the $(t_{2g})^{4}$ 
sector and in one dimension. Fixing the Hund coupling to a robust
value compatible with some multiorbital materials, we present the phase diagram varying 
the Hubbard $U$ and spin-orbit coupling $\lambda$, at zero temperature.
Our results are shown to be qualitatively similar to those recently reported using the Dynamical 
Mean Field Theory in higher dimensions, providing a robust basis to approximate many-body
techniques. Among many results, we observe an interesting transition from an orbital-selective 
Mott phase to an excitonic insulator with increasing $\lambda$ at intermediate $U$.
In the strong $U$ coupling limit, we find a non-magnetic insulator with an effective angular momentum 
$\langle(\bold{J}^{eff})^{2}\rangle \ne 0$ near the excitonic phase, smoothly 
connected to the $\langle(\bold{J}^{eff})^{2}\rangle = 0$ regime.
We also provide a list of quasi-one dimensional materials where the physics discussed in this publication could be realized.

\end{abstract}


\maketitle

\section{Introduction}

The study of iridates continues attracting considerable attention. In layered materials such
as Sr$_2$IrO$_4$ and Ca$_2$IrO$_4$, involving $5d$ electrons, the Hubbard 
repulsion is moderate as compared to
$3d$ electrons because the size of the associated wave functions is larger 
for the $5d$ sector~\cite{iri1,iri2,iri3,iri4,iri5,iri6,Rau01,Haskel01}.
In addition, as we move down in the periodic table the strength of the spin-orbit coupling (SOC)
increases as $Z^4$, with $Z$ as the atomic number, and it can become of order 0.4~eV for some
$4d$ or $5d$ materials. As a consequence, iridates provide an interesting playground where 
the Hubbard repulsion and SOC are of similar magnitudes~\cite{Cao03}.
In  these iridates  the $t_{2g}$ orbitals split into a total angular
momentum (half-filled) doublet $j=1/2$ and a (fully occupied) quartet $j=3/2$~\cite{BJKim01}.

More recently, interest also developed in other transition metal oxides with octahedron 
or distorted octahedron crystal-field splittings leading
to $(t_{2g})^{4}$ ions~\cite{BJKim01,4d-1,4d-2,4d-4,4d-5,4d-6,4d-7,4d-8,Phelan01,Cao01,Terizc01,TDey01,Corredor01,Meetei01,Svoboda01}. 
When the Hubbard $U$ and Hund $J_{H}$ couplings are large it is expected that the system  
develops $S=1$ states, while increasing the SOC $\lambda$ should lead to states with an effective angular
momentum zero. Thus, the next obvious step is to understand the phases in these systems in the presence 
of hopping. Experiments on these materials have shown contrasting results thus far. For example, 
the magnetic properties of Sr$_{2}$YIrO$_{6}$~\cite{Cao01} suggest 
exotic antiferromagnetic (AFM) ordering coming from excitonic condensation, while other 
experiments~\cite{Corredor01} favor a nonmagnetic ground state. Double perovskites such as 
Ba$_{2}$YIrO$_{6}$ are also challenging to study~\cite{Terizc01,TDey01}. 
This situation demands a comprehensive and accurate theoretical study of the combined 
effects of $U$ and $\lambda$ in the $(t_{2g})^{4}$ sector.

Alongside the iridates, progress has been made on iron-based superconductors in recent years~\cite{johnston,peter,scalapino}.
While initially the expectation was that weak coupling approximations and Fermi surface nesting 
between hole and electron pockets could be sufficient to 
understand these compounds, recent efforts have highlighted the importance of Hubbard interactions
of at least intermediate value between weak and strong coupling~\cite{daiNatPhys}. For example, there are
materials that do not have hole pockets, yet they still superconduct~\cite{SCnohpockets}. 
Moreover, via angle-resolved photoemission spectroscopy (ARPES) it has been argued~\cite{Borisenko2016}
that a SOC of order 20~meV, much smaller than in iridates, 
may still influence the features of the Fermi level and thus affect superconducting properties.

Considering all these challenging fields of research, and their common focus on intermediate range Hubbard $U$
and spin-orbit coupling $\lambda$ interactions, in this publication employing numerically exact computational
techniques we will study a model of interacting electrons in the simultaneous presence of nonzero $U$, $\lambda$,
and $J_H$. In particular, we will analyze a multiorbital model defined on a one dimensional geometry.

Our study is conceptually generic but for simplicity will focus on a previously used
three-orbital Hubbard model with bands that resemble layered iron superconductors, 
containing hole and electron pockets. In the absence 
of spin-orbit coupling, this model was studied before via the Density Matrix Renormalization 
Group (DMRG) technique 
and a rich phase diagram was observed, including an orbital-selective 
Mott phase (OSMP), where two orbitals are partially filled and thus they are metallic, 
while the other orbital is half-filled and behaves 
like a Mott insulator~\cite{Julian01,Julian02,Guangkun01,Shaozhi01}. Our main focus is to analyze how this phase diagram is modified after including atomic spin-orbit effects. 
The generic analysis reported here is important for three reasons:

{\it (i)} By constructing the phase diagram in one dimension including the combined effects of the Hubbard interaction $U$ as well 
as the spin-orbit coupling $\lambda$ with a  robust 
computational technique, we can address the accuracy of previous 
approximate studies performed in higher dimensions. 
For example, recently Dynamical Mean Field Theory (DMFT) 
calculations were performed~\cite{Sato01,Sato02,JKim01bis} on a  
three-orbital Hubbard model with degenerate $t_{2g}$ orbitals 
and four electrons per site. Their analysis showed the presence 
in the phase diagram of an interesting excitonic condensate (to 
be described below) and a non-magnetic insulator with zero effective 
total angular momentum. Our accurate numerical results on chains 
using non-cubic $t_{2g}$ bands confirm most of the DMFT predictions, 
including the existence of an excitonic condensate,
thus suggesting that studies in different dimensions may lead to qualitatively similar results.

{\it (ii)} There are real materials with quasi-one dimensional characteristics where spin-orbit effects are expected to be important. 
For example, recently, single-crystals of Ba$_5$AlIr$_2$O$_{11}$ that contain 
dimer chains were experimentally studied~\cite{Terzic02}. 
This is a Mott insulator with a subtle structural transition at $T_S = 210$~K and a magnetic transition at much lower temperatures. 
A novel and intriguing magnetic state was reported, that is neither $S=3/2$ nor $J=1/2$ but instead intermediate between them. 
Other examples of spin-chain 4$d$- and 5$d$-based compounds are Sr$_5$Rh$_4$O$_{12}$, Ca$_5$Ir$_3$O$_{12}$, and Ca$_4$IrO$_6$~\cite{Cao02}. 
These are insulators characterized by partial AFM order at low temperatures. Sr$_3$CuIrO$_6$ is also a quasi-one dimensional 
material where IrO$_6$ octahedra are linked by spin-1/2 Cu ions along one direction~\cite{XLiu01}. In this compound intersite hopping is 
suppressed by the geometry of the system locating Sr$_3$CuIrO$_6$ in the strongly localized regime, with a noncubic crystal-field comparable 
in strength to the spin-orbit coupling. Other examples of interesting one dimensional systems where our results may be of relevance 
are BaIrO$_3$~\cite{cao2000,Marco01}, CaIrO$_3$~\cite{Bogdanov01}, Sr$_3$$M$IrO$_6$ ($M$ = Ni, Cu, Zn)~\cite{nguyen1995}, lead iodides~\cite{xiong2015}, 
and alkaline-earth palladates~\cite{wang1999}.

{\it (iii)} As already explained, recent ARPES measurements reported 
a sizable spin-orbit splitting 
in all the main members of the iron-based superconductors family~\cite{Borisenko2016}. This spin-orbit coupling affects the low-energy electronic 
structure and, thus, may have implications for superconductivity. While the magnitude of $\lambda$ for iron pnictides and chalcogenides 
is substantially smaller than for iridates, it is conceptually interesting to investigate what kind of phases could be found 
if members of the iron superconductors family would have a larger $\lambda$.

Spin-orbit effects are often discarded in the literature, usually
by hand waving arguments, and the models are largely simplified as a result. 
But realistic detailed studies involving spin-orbit couplings comparable to other small 
energies of interest (such as the magnetic superexchange $J$) are lacking. 
Moreover, it is experimentally challenging to determine the precise magnitude of spin-orbit contributions. 
On the $ab$-$initio$ side of theory, often these spin-orbit 
contributions are not considered if expected to be smaller than systematic 
errors in the approach, typically of order 0.5~eV. As a 
consequence, an evaluation of the effects of spin-orbit corrections 
on the results of specific models could determine if refined $ab$-$initio$ or measurements are required.

The organization of this manuscript is as follows. In Sec.~II, the model used and the computational methodology are presented. 
In Sec.~III, the main results, particularly the phase diagram varying $U$ and
$\lambda$, are shown. In particular, we address three regimes: weak, intermediate, and strong Hubbard interaction $U$. 
In Sec.~IV, we discuss the results and present our conclusions. 

\section{Model and Method}\label{models}

In this study we have used a one dimensional three-orbital Hubbard model. The Hamiltonian contains a tight-binding term, 
an on-site Hubbard interaction, and a spin-orbit coupling: $H = H_{K} + H_{\mathrm{int}} + H_{SOC}$. The electronic kinetic energy component is 
\begin{equation}
H_{K} = -\sum_{{i},\sigma,\gamma,\gamma^{\prime}}t_{\gamma\gamma^{\prime}}
(c_{{i}\sigma\gamma}^{\dagger}c^\pdag_{{i}+1\sigma\gamma^{\prime}}+\mathrm{h.c.})
+\sum_{{i},\sigma,\gamma}\Delta_{\gamma}n_{{i}\sigma\gamma}.
\end{equation}
The hopping amplitudes $t_{\gamma \gamma'}$ are defined in orbital space and they connect the nearest-neighbor lattice sites ${i}$ and ${i}+1$, with the 
specific values (in eV units) $t_{00}=t_{11}=-0.5$, $t_{22}=-0.15$, and $t_{\gamma\gamma^{'}}=0$ if $\gamma\ne \gamma^{'}$. 
The total bandwidth is $W=4.33\, |t_{00}|$. The above mentioned 0, 1, and 2 orbitals 
can be visualized as representing  the canonical $d_{yz}$,   $d_{xz}$, and  $d_{xy}$ orbitals, 
respectively. The orbital-dependent crystal-field splitting is denoted by $\Delta_\gamma$, 
with $\Delta_0 = -0.05$, $\Delta_1 = -0.05$, and $\Delta_2 = 0.8$ (also in eV units). 
The band structure of this model qualitatively resembles that of iron-based superconductors, i.e., hole and electron pockets 
centered at wavevectors $q = 0$ and $\pi$, respectively. A very similar band structure 
was used in our previous studies for a three-orbital 
Hubbard model~\cite{Julian01,Julian02,Guangkun01,Shaozhi01}, where OSMP was analyzed. This previous work was carried out 
in the absence of spin-orbit interactions, and our main focus is to analyze the effects of this additional term in the model. 
The Hubbard portion of the Hamiltonian includes the following onsite components in the standard notation
\begin{multline}\label{INT_term}
H_{\mathrm{int}} = U\sum_{{i},\gamma} n_{{i}\uparrow\gamma}
n_{{i}\downarrow\gamma} 
+\left(U'-J_{H}/2\right)\sum_{{i},\gamma<\gamma'} n_{{i}\gamma}
n_{{i}\gamma'} 
\\
  -2J_{H}\sum_{{i},\gamma<\gamma'} \mathbf{S}_{{i}\gamma} \cdot 
  \mathbf{S}_{{i}\gamma'} 
+J_{H}\sum_{{i},\gamma<\gamma'} \left( P^{\dagger}_{{i}\gamma} 
P_{{i}\gamma'} + \mathrm{h.c.} \right) . 
\end{multline}
In this expression the operator $\mathbf{S}_{{i}\gamma}= {{1}\over{2}}\sum_{\alpha,\beta} 
c_{{i}\alpha\gamma}^{\dagger} \sigma_{\alpha\beta} c^\pdag_{{i}\beta\gamma}$ is 
the total spin  at orbital $\gamma$ and lattice site ${i}$, and $n_{{i}\gamma}$ is the electronic 
density at each orbital. The first two terms describe the intra- and inter-orbital electronic repulsion, respectively. 
The third term contains the Hund coupling that favors the ferromagnetic alignment of the spins at different orbitals; the fourth term is the 
pair hopping with $P_{{i}\gamma}=c_{{i}\downarrow\gamma}c_{{i}\uparrow\gamma}$ 
as the pair operator. We use the standard relation $U^{\prime}=U-2J_H$ based on rotational invariance, and we fix $J_{H}=U/4$ because this value is
widely accepted in iron superconductors to be realistic~\cite{daiNatPhys}. For these reasons, only $U$ and $\lambda$ are free parameters in our
study. Future work can analyze in more detail the influence of varying the Hund coupling as well as other parameters in the model. 

The SOC term is 
\begin{equation}\label{SO_term}
H_{\mathrm{SOC}}=\lambda\sum_{{i},\gamma,\gamma^{'},\sigma,\sigma^{'}}
{{\langle \gamma|{\bold{L}_{i}}|\gamma^{'}\rangle}\cdot{\langle\sigma|{\bold{S}_{i}}|\sigma^{'}\rangle}}
c_{i\sigma\gamma}^{\dagger}c_{i\sigma^{'}\gamma^{'}} \hspace{0.1cm},
\end{equation} 
where $\lambda$ is the SOC coupling strength, as already explained. 
Because of the presence of the SOC term 
the total spin along the $z$-axis, $S_{z}$, is no longer 
a good quantum number; hence, we cannot target specific $S_{z}$ sectors 
in our numerical DMRG calculation. To reduce the computational cost, we have instead selected the parameters contained 
in $H_{K}$ such that $[H,J^{eff}_{z}]=0$ 
where ${\bold{J}}^{eff}=\sum_{i}({\bold{S}_{i}} - {\bold{L}_{i}})$ 
. Note that for arbitrary values of the hopping amplitudes and crystal-fields, 
$J^{eff}_{z}$ is also not a good quantum number as discussed in the Appendix~\ref{appendix1}. We then target subspaces 
with a fixed total $J^{eff}_{z}=\sum_{{i}}(J^{eff}_{z})_i$ for the system. 
The SOC term is diagonalized in the $j^{eff}$ basis, where $j^{eff}$ 
is the quantum number associated with $\bold{J}^{eff}$ (to avoid complications 
in the notation, as when $j^{eff}$ should appear as subindex, 
sometimes this quantum number will be denoted simply by $j$).
$m$ is the projection along the $z$-axis namely 
the quantum number of ${J}^{eff}_{z}$ . The fact that the 
good quantum numbers for the SOC term are associated with the effective angular momentum, instead of the total angular 
momentum (${\bold{J}}={\bold{S}} + {\bold{L}}$), is a consequence of the ``$t_{2g} - p$'' 
equivalence discussed in~\cite{AbragamAndBleaney}. The ``$t_{2g}$ subspace'' of the $d$-orbitals 
($l=2$ for a complete $d$ orbital set) has $\langle{\bf{L}}^{2}\rangle$=2 
for a single electron, hence ``$t_{2g}$'' 
is isomorphic to the $l=1$ space (i.e., the $p$-orbitals) 
under the following mapping: $|1\rangle_{p} \equiv-i|-1\rangle_{d}$, 
$|-1\rangle_{p} \equiv i|1\rangle_{d}$, 
$|0\rangle_{p} \equiv  |xy\rangle_{d}$, 
and $\bold{L}^{l=1}$ $\equiv$ -$\bold{L}^{t_{2g}}$. 

\begin{figure}[!th]
\hspace*{-0.1cm}
\vspace*{-0.3cm}
\begin{overpic}[width=\columnwidth]{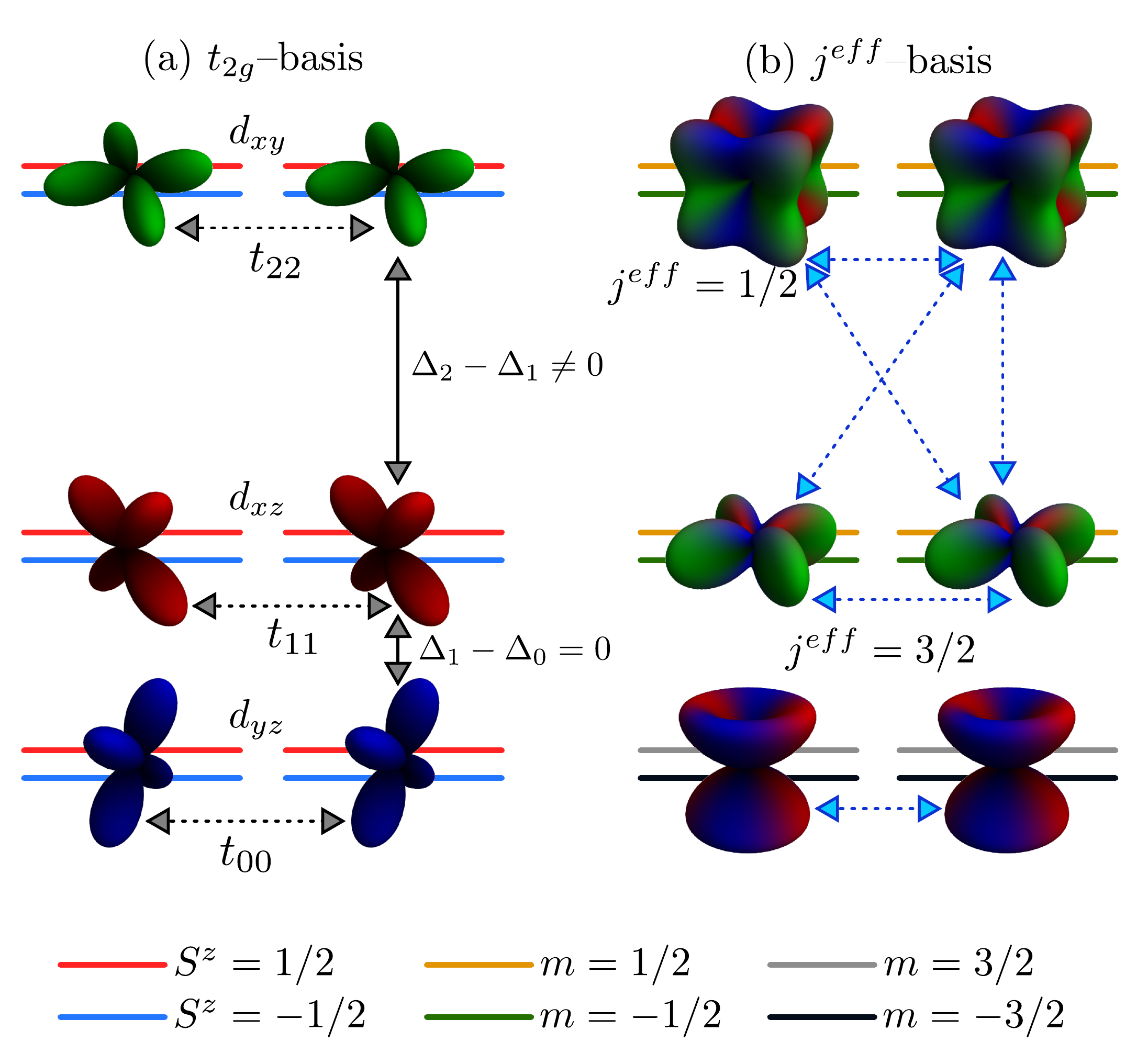}
\end{overpic}
\caption{In panel (a) we show the connections between $t_{2g}$ orbitals using dashed lines, while in panel (b) the dashed lines represent the non-zero 
connections present in the ($j^{eff},m$) basis if we use the proper hopping and crystal-field parameters 
satisfying the constraints described in the Appendix~\ref{appendix1}.}
\label{fig1}
\end{figure}

The transformation between the $t_{2g}$ orbitals and the $j^{eff}$ basis is given by (dropping site $i$ index)
\begin{equation}\label{transformation}
\renewcommand{\arraystretch}{1.5}
\begin{bmatrix}a_{\frac{3}{2},\frac{3s}{2}}\\a_{\frac{3}{2},-\frac{s}{2}}\\a_{\frac{1}{2},-\frac{s}{2}}\end{bmatrix}
= \begin{bmatrix}\frac{is}{\sqrt{2}}&\frac{1}{\sqrt{2}}&0\\\frac{s}{\sqrt{6}}&\frac{i}{\sqrt{6}}&\frac{2}{\sqrt{6}}\\
\frac{-s}{\sqrt{3}}&\frac{-i}{\sqrt{3}}&\frac{1}{\sqrt{3}}
\end{bmatrix}\begin{bmatrix}c_{\sigma yz}\\c_{\sigma xz}\\c_{\bar{\sigma} xy}\end{bmatrix} ,
\end{equation}
where $s$ is $1(-1)$ when $\sigma$ is $\uparrow(\downarrow)$ and $\bar{\sigma}=-\sigma$. The $H_{SOC}$ term in the $j^{eff}$ basis becomes
\begin{eqnarray}\label{H_SO}
H_{\mathrm{SOC}}&=&\sum_{{i}}\frac{\lambda}{2}(-a_{{i},\frac{3}{2},\frac{3}{2}}^{\dagger} a_{{i},\frac{3}{2},\frac{3}{2}}^{\phantom{\dagger}}
- a_{{i},\frac{3}{2},-\frac{1}{2}}^{\dagger} a_{{i},\frac{3}{2},-\frac{1}{2}}^{\phantom{\dagger}}
\nonumber\\
&-&a_{{i},\frac{3}{2},-\frac{3}{2}}^{\dagger} a_{{i},\frac{3}{2},-\frac{3}{2}}^{\phantom{\dagger}}
-a_{{i},\frac{3}{2},\frac{1}{2}}^{\dagger} a_{{i},\frac{3}{2},\frac{1}{2}}^{\phantom{\dagger}}
\nonumber\\
&+&2a_{{i},\frac{1}{2},\frac{1}{2}}^{\dagger} a_{{i},\frac{1}{2},\frac{1}{2}}^{\phantom{\dagger}} 
+ 2a_{{i},\frac{1}{2},-\frac{1}{2}}^{\dagger} a_{{i},\frac{1}{2},-\frac{1}{2}}^{\phantom{\dagger}})\,.
\end{eqnarray}
The SOC component commutes with ${(\bold{J}^{eff})}^2$. As a consequence, in the $H_{\mathrm{SOC}}$ term 
there is four(two)-fold degeneracy in the $j^{eff}={3}/{2}$ (${1}/{2}$) bands. However, the four-fold degeneracy 
of the $j^{eff}={3}/{2}$ sector breaks into a pair of two-fold Kramer degeneracies due to the presence of the non-cubic $t_{2g}$-band structure 
used in our model. This can be understood by analyzing the $H_K$ term in the ($j^{eff},m$) basis. 
In Fig.~\ref{fig1} we show explicitly the connections contained in $H_{K}$ between the $t_{2g}$ states and the corresponding connections 
between $(j^{eff},m)$ states, after imposing the constraints on the hopping and crystal-field parameters (see Appendix~\ref{appendix1}). 
We have noticed that the non-cubic nature of  the $t_{2g}$ states (i.e., the non-degeneracy of the $d_{xy}$ with the \{$d_{xz}$, $d_{yz}$\} states, 
consequence of the tetragonal type $t_{2g}$ bands) leads to hybridization between $(j^{eff}=1/2,m=\pm1/2)$ and $(j^{eff}=3/2,m=\pm1/2)$ states. 
This hybridization breaks the four-fold degeneracy of the $j^{eff}=3/2$ states and also leads to the formation of new bands in which $H_{K} + H_{SOC}$ is diagonalized.

After using the inverse transformation of Eq.~(\ref{transformation}) in the tight-binding term, we diagonalized the $H_{K} + H_{SOC}$ together to obtain the following bands
\begin{equation}\label{bands}
H_{K} + H_{SOC}=\sum_{k,\alpha,s}E_{\alpha}(k)\tilde{a}^{\dagger}_{k,\alpha,s}\tilde{a}_{k,\alpha,s}\,,
\end{equation}
where $s \in {\{1,-1\}}$ and $\alpha \in {\{0,1,2\}}$. Here $\alpha$ is the band index, and the relation 
between $\tilde{a}^{\dagger}_{k,\alpha,s}$ and $a_{k,j,m}$ is shown in the Appendix~\ref{appB}. The dispersion 
relations for the bands are $E_{0}(k)=\epsilon_{0}(k)-\frac{\lambda}{2}$, and $E_{\alpha}(k)=\frac{1}{2}[\epsilon_{2}(k) + \epsilon_{1}(k) + \frac{\lambda}{2} + (-1)^{\alpha}\sqrt{ (\epsilon_{2}(k) - \epsilon_{1}(k) -\frac{\lambda}{2})^{2} + 2\lambda^{2}}]$ 
for $\alpha \in \{1,2\}$; where $\epsilon_{\alpha}(k)=-2t_{\alpha\alpha}\cos(k)+\Delta_{\alpha}$ for $\alpha \in \{0,1,2\}$. 
At $\lambda=0$, the bands 0, 1, and 2 reduce to the standard bands of the $d_{yz}$, $d_{xz}$, and $d_{xy}$  orbitals, respectively. 
For $\lambda/W \gg 0$, the bands 1 and 2 reduce to the $(j^{eff}=1/2,m=\pm1/2)$ and $(3/2,\pm1/2)$  
states, respectively, and $n_{3/2,\pm3/2}=\tilde{n}_{0\pm1}$ for any $\lambda$. The above described 
non-interacting portion of the Hamiltonian is useful to understand the effect of spin-orbit coupling 
in the small $U/W$ region of the phase diagram, as discussed below.

Our many-body calculations are performed using the DMRG technique~\cite{White01,White02,White03}
 applied to one dimensional chains of various system lengths, such as $L$ = 8, 16, 24, and 32 sites. 
We have used up to 600 states for the DMRG process and have maintained a truncation 
error below $10^{-14}$ throughout the finite 
algorithm sweeps. In the latter, we performed 10 to 15 full sweeps to gain 
convergence depending on the system size.  
We studied the presence of various phases by calculating expectation values of 
$n_{{i}\alpha}$, $n_{{i}jm}$, $\bold{S}^{2}_{{i}}$, $\bold{L}^{2}_{{i}}$, 
$(\bold{J}^{eff})_{i}^{2}$ , the canonical spin structure factor $S(q)$, and the exciton
pair-pair correlation $\langle \Delta_{jm}^{\dagger\tilde{j}m}(i)
\Delta_{jm}^{\tilde{j}m}(i^{'}) \rangle$ (defined in III.B).

\begin{figure}[!t]
\hspace*{-0.52cm}
\vspace*{0cm}
\begin{overpic}[width=1.1\columnwidth]{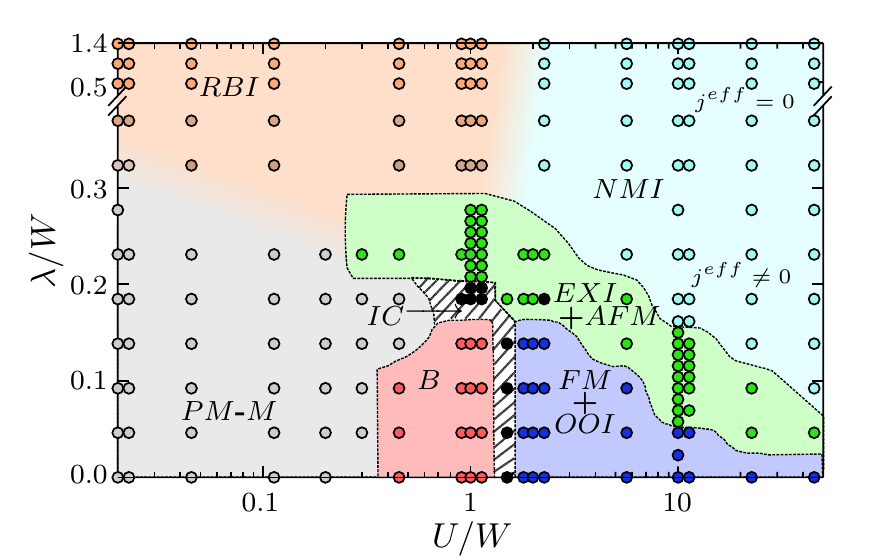}
\end{overpic}
\caption{$\lambda$-$U$ phase diagram (note the log scale in $U/W$-axis). RBI, PM-M, B, FM, OO, IC, EXI, AFM, and NMI stands for relativistic band insulator, 
paramagnetic metal, block phase, ferromagnetic, orbital ordering, incommensurate, 
excitonic insulator, antiferromagnetic, and nonmagnetic insulator, respectively. 
Lines separating phases are guides to the eyes. The actual
small circles indicate specific values of data 
points that were investigated with DMRG. Their high density indicates that this
effort has been computationally demanding.}
\label{fig2}
\end{figure}

\section{Results}\label{results}

The main result of this publication, presented in Fig.~\ref{fig2}, is the phase diagram of the three-orbital Hubbard model analyzed here, varying
$U$ and $\lambda$ in units of the bandwidth $W$ at a fixed electronic density of four electrons per site on average. In the following subsections, details are provided for the three special cases
of weak, intermediate, and strong Hubbard $U$ coupling. Also note that our study is in one dimension
and for this reason when we write that at some values of $U$ and $\lambda$ we are at a 
phase with some particular characteristics, this has to be interpreted in the sense of 
dominant power-law decaying correlations as opposed to true long-range order.

\subsection{Paramagnetic Metal and Relativistic Band Insulator (Weak Coupling)}\label{Paramagnetic metal and Relativistic band insulator}
\begin{figure}[]
\hspace*{-0.0cm}
\begin{overpic}[width=\columnwidth]{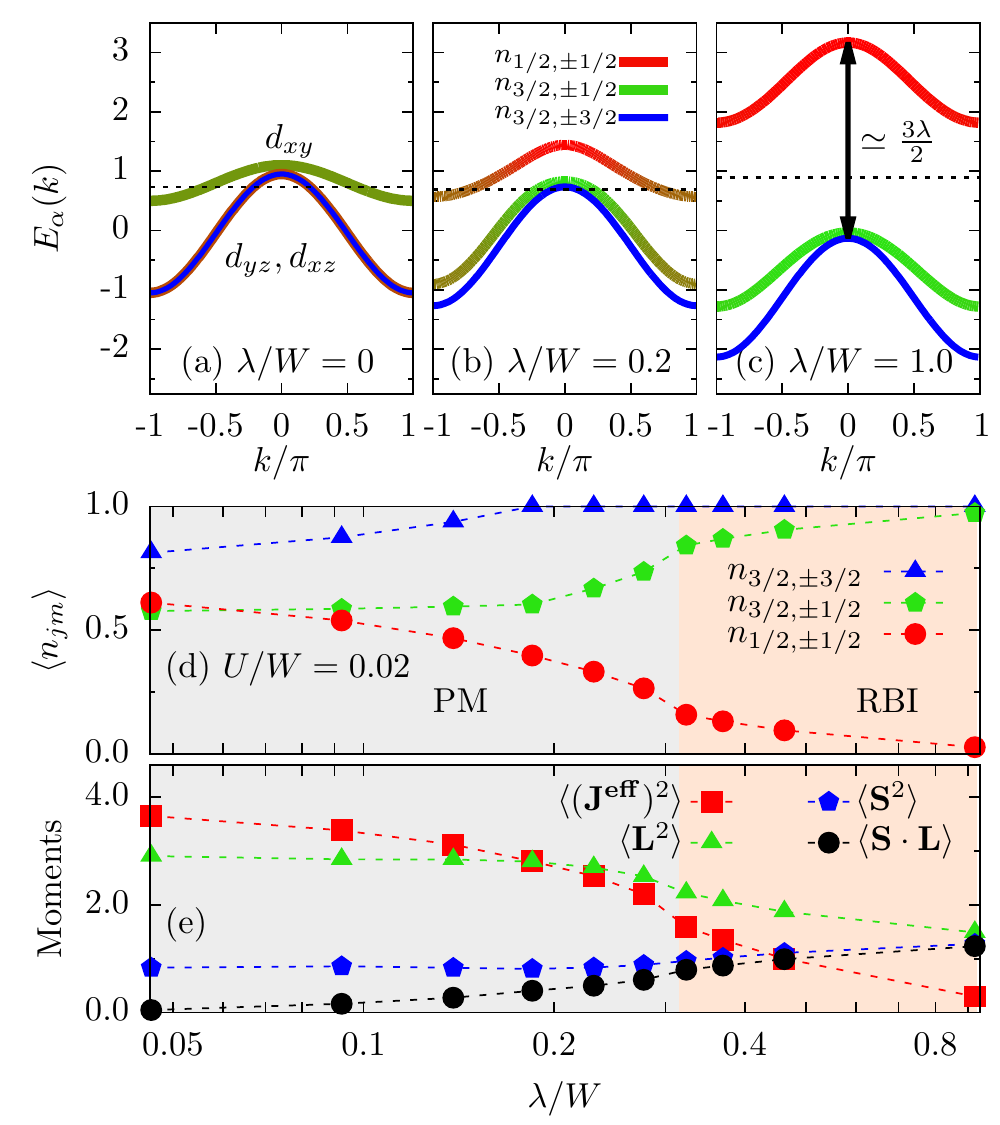}
\end{overpic}
\caption{Panel (a) shows the non-interacting bands of our model at $\lambda/W=0.0$. As explained in the
text, the almost fully populated bands are degenerate and superimposed. In (b) and (c),
we show the bands at ${\lambda}/W=0.2$ and 1.0, respectively. Panel (c) displays a clear
opening of a gap, i.e., the system becomes a band insulator. Colors are decided depending on the relative
contributions from the three $(j^{eff},m)$ bands, with the pure cases shown in the legend of panel (c). 
Panel (d) contains the occupation numbers in the $(j^{eff},m)$ basis, while
panel (e) has the local magnetic moments strengths (see legend) as well as $\langle{\textbf{S}\cdot\textbf{L}}\rangle$, all at $U/W=0.02$. Calculations for panels (d) and (e) were performed with DMRG using a 
$L=16$ chain, while panels (a,b,c) are from exact analytical formulas.}
\label{fig3}
\end{figure}

First, we will briefly discuss the small $U$ region, i.e., the weak coupling limit. 
This regime can be understood by
analyzing the non-interacting limit using Eq.~(\ref{bands}). Varying the strength of the spin-orbit
coupling $\lambda$ at $U/W=0$ the exact band structure is shown in Fig.~\ref{fig3}(a,b,c). From this
analysis we expect the presence of a trivial paramagnetic metal (PM-M) at small $\lambda$
which transforms into the relativistic band insulator (RBI) 
regime by increasing $\lambda$. At $U=0$, for four electrons per site, we can use the 
condition $E_{2}(k=\pi)=E_{1}(0)$ to calculate analytically the critical spin-orbit coupling strength $\lambda_{c}$ for which a gap opens:
\begin{widetext}
\begin{equation}
\lambda_{c}(U=0)=\frac{2t_{11}t_{22}(\Delta_{2}-\Delta_{1}) +
2(t_{11} + t_{22}) \sqrt{t_{11}t_{22}(4t_{11}t_{22} + 
8(t_{11}+t_{22})^{2}-2(\Delta_{2}-\Delta_{1})^{2})}}{2(t_{11}+t_{22})^{2}+t_{11}t_{22}}\,.
\end{equation}   
\end{widetext}
The value of $\lambda_{c}/W$ for our specific hopping parameters and crystal-field splittings 
is $\simeq 0.33$. The state $(j^{eff}=3/2,m=\pm3/2)$ moves below the Fermi level before  
$\lambda$ approaches $\lambda_{c}$ as $(j^{eff}=3/2,m=\pm3/2)$ does not hybridize with any other state.
For the $U \ne 0$ case, but still small, $\lambda_{c}$ can be different from $\lambda_{c}(U=0)$. 
We suspect $\lambda_{c}$ decreases monotonically as $U$ increases because at intermediate 
$U$ the excitonic insulator regime develops (see Sec.~III.B) 
for $\lambda$ lower than $\lambda_c(U=0)$, and this Bardeen-Cooper-Schrieffer (BCS)
limit of the excitonic insulator (EXI) regime (discussed in next Section) at intermediate $U$ should be present near the semimetal-semiconductor transition
as discussed before~\cite{rice1960}. 
The decrease in $\lambda_{c}$ is a result of renormalization of bands due to 
correlation effects, which enhances the effect of spin-orbit coupling, as discussed in~\cite{LDu01}.
In Fig.~\ref{fig3}(d), we
show the occupation of the $(j^{eff},m)$ bands ($n_{jm}$ are the respective densities) varying
$\lambda/W$ at a fixed $U/W=0.02$, displaying a smooth crossover from paramagnetic
metal to band insulator. At large $\lambda$ the $j^{eff}={3}/{2}$ bands are completely
filled, while the $j^{eff}$=${1}/{2}$ band is nearly empty at $\lambda/W=1.0$ and its
population continues decreasing as $\lambda$ is further increased.

In contrast to previous DMFT studies performed for three degenerate
bands~\cite{Sato01,Sato02}, the four-fold degeneracy of the $j^{eff}=3/2$ bands is here
explicitly broken due to the hybridization between the (${3}/{2}, \pm{1}/{2}$) and
(${1}/{2}, \pm{1}/{2}$) states. This is a consequence of a non-cubic crystal-field splitting
and specific hopping parameters to resemble iron-based
superconductors, as explained before. We also observed the above
mentioned splitting in the intermediate and strong Hubbard coupling limits, thus, 
this effect propagates into the interacting region. It is important to
mention here that due to the hybridization of our model, in the RBI regime the $j^{eff}=1/2$
state can have a non-zero occupation because it can have non-zero weight in the band below
the Fermi surface. In other words, due to the hybridization between the (${3}/{2}, \pm
{1}/{2}$) and (${1}/{2}, \pm{1}/{2}$) states, the basis where $H_{K} + H_{SOC}$ is
diagonalized corresponds to $\tilde{a}_{{k},\alpha,{s}}$, not ${a}_{{k},j,{m}}$. As a consequence, in
the lower portion of the RBI region in the phase diagram we have a finite occupation of the
(${1}/{2}, \pm{1}/{2}$) states coexisting with a sharp band insulator gap at the Fermi
level. Only as the spin-orbit coupling continues increasing is that $\tilde{a}_{{k},\alpha,{s}}$
reduces asymptotically to ${a}_{{k},{j,m}}$, and we reach zero occupation of the
(${1}/{2},\pm{1}/{2}$) states.\par

Note that a similar splitting between the $j^{eff}={3}/{2}$, $m=\pm{1}/{2}$ and
$j^{eff}={3}/{2}$, $m$=$\pm{3}/{2}$ bands of nearly 0.7~eV has also been observed
in the $(t_{2g})^{5}$ perovskite CaIrO$_{3}$~\cite{Bogdanov01} as a result of the presence of a
non-cubic crystal-field, although our study is not directly related to this material.\par

Figure~\ref{fig3}(e) shows the local moments $\langle (\textbf{J}^{eff})^{2} \rangle$, 
$\langle \textbf{L}^{2} \rangle$, and
$\langle \textbf{S}^{2} \rangle$, as well as $\langle{\bf S\cdot L}\rangle$. Similarly to the non-interacting case, at
$U/W$=0.02, the moments $\langle \textbf{L}^{2} \rangle$, $\langle \textbf{S}^{2} \rangle$, 
and $\langle{\bf S\cdot L}\rangle$
converge to $4/3$ while $\langle (\textbf{J}^{eff})^{2} \rangle$ 
tends to 0 for large spin-orbit coupling (this can
be checked by using the atomic state 
$a_{\frac{3}{2},\frac{3}{2}}^{\dagger}a_{\frac{3}{2},-\frac{3}{2}}^{\dagger}a_{\frac{3}{2},\frac{1}{2}}^{\dagger}
a_{\frac{3}{2},-\frac{1}{2}}^{\dagger}|0\rangle$, which is the ground state of the
$H_{SOC}$ term).

\subsection{Excitonic Insulator and Orbital Selective Mott Phase (Intermediate Coupling)}

\begin{figure}[!th]
\hspace*{-0.0cm}
\begin{overpic}[width=\columnwidth]{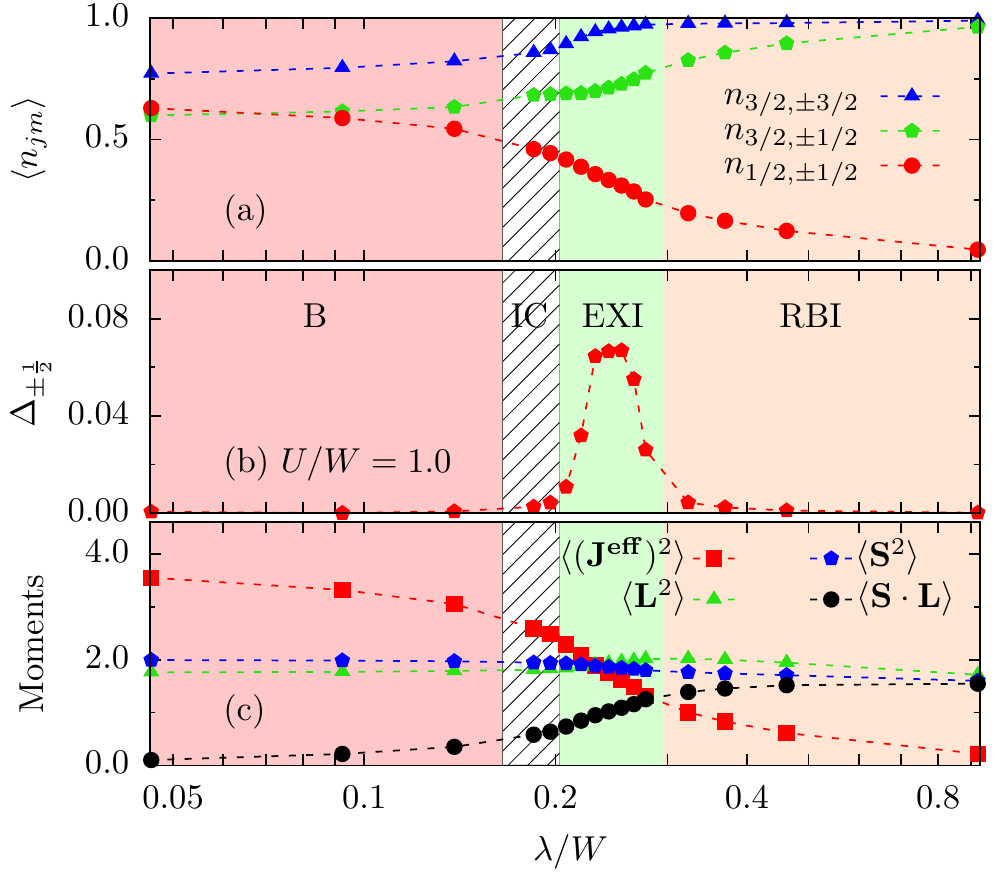}
\end{overpic}
\caption{DMRG results obtained at $U/W=1$ (intermediate coupling) and using a $L=16$ system.  
Panel (a) shows occupation number in the $(j^{eff},m)$ bands while (b) shows the excitonic parameter $\Delta_m$ defined in Eq.(8) 
varying $\lambda/W$. Panel (c) shows the three local moment strengths as well as $\langle{\textbf{S.L}}\rangle$. }
\label{fig4}
\end{figure}

\begin{figure}[!th]
\hspace*{-0.04cm}
\begin{overpic}[width=\columnwidth]{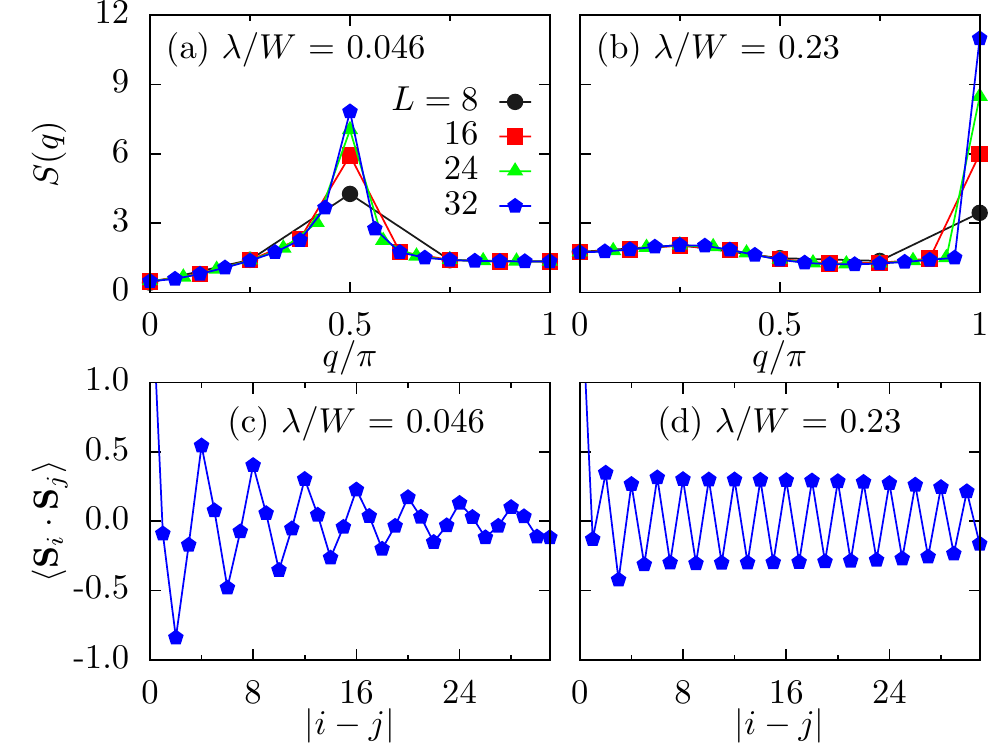} 
\end{overpic}
\caption{DMRG resuls obtained at $U/W=1$. Panels (a) and (b) contain the spin structure factor in the block phase and in the
EXI phase, respectively, for various number of sites $L=8$ (black), 16 (red), 24 (green), and 32 (blue).
Panels (c) and (d) display the real-space spin-correlations at $L=32$ 
corresponding to the block and EXI phases, respectively, for the $\lambda/W$'s indicated. }
\label{fig5}
\end{figure}

In this subsection we will discuss the results obtained at intermediate Hubbard interaction. 
This region is difficult and it cannot be treated 
perturbatively, thus numerical exact studies via the DMRG method are important. 
In this regime we have found several interesting phases such as the OSMP, EXI, incommensurate phase, 
and at large $\lambda/W$ we again found the RBI of weak coupling. In Fig.~\ref{fig4}, we present results obtained at  $U/W=1.0$. 
At small $\lambda$, we reproduced the OSMP regime with a magnetic Block arrangement of the spins 
($\uparrow \uparrow \downarrow \downarrow\uparrow \uparrow$)~\cite{Julian01,Julian02,Guangkun01,Shaozhi01}. 
The presence of OSMP features is confirmed by measuring the occupation 
of the $t_{2g}$ states: in this regime the $d_{xy}$ orbital has occupation very close to 1, 
while $d_{xz(yz)}$ has occupation nearly 1.5 (see Appendix~C). 
The spin structure factor $S(q)$ and the real-space spin-spin correlations are shown in Fig.~\ref{fig5}(a,c) 
at $\lambda/W$=$0.046$  providing evidence for the Block magnetic order.

Figure~\ref{fig4}(a) shows the occupation number 
in the ($j^{eff},m$) states corresponding to $U/W$=$1.0$ at different $\lambda$'s. 
As in the case of weak coupling, here the system also converges to a band insulator 
at sufficiently large spin-orbit coupling as the $j^{eff}=3/2$ state is completely 
filled and $j^{eff}=1/2$ becomes empty. In the strength of the magnetic moments 
we have noticed a clear difference between the intermediate and weak coupling regimes, 
as shown Fig.~\ref{fig4}(c). We found $\langle{\bf{S}}^{2}\rangle=2$ 
in the OSMP and in the incommensurate phase. However, this quantity is reduced 
after entering in the EXI phase, and at the same time $\langle{\bf{L}}^{2}\rangle$ increases. 
We also noticed that 
for any Hubbard interaction in the limit of sufficiently large $\lambda$, $\langle{\bf{S}}^{2}\rangle$=$\langle{\bf{L}}^{2}\rangle$=$\langle{\bf{S\cdot L}}\rangle$ 
which means $\bf{S}$ and $\bf{L}$ become parallel to each other. 
As a consequence, $\langle{(\bf{J}}^{eff})^{2}\rangle=\langle{\bf{S}}^{2}\rangle + \langle{\bf{L}}^{2}\rangle - 2\langle{\bf{S\cdot L}}\rangle$ converges to 0.

To identify the EXI phase, we calculated a pair-pair correlation function (note, here ``pair'' denotes an {\it{electron-hole}} pair), i.e., $\langle \Delta_{jm}^{\dagger\tilde{j}m}(i) \Delta_{jm}^{\tilde{j}m}(i^{'}) \rangle$, 
where $\Delta_{jm}^{\tilde{j}m}(i) 
= a_{{i} \tilde{j}m}^{\dagger} a_{{{i}}jm}^{\phantom{\dagger}}$
(here we fixed $j=1/2$ and $\tilde{j}={3/2}$). 
This operator was already introduced in previous literature~\cite{Sato01,Sato02}.
In our DMRG calculations, and in agreement with~\cite{Sato01}, 
we noticed that in the EXI
phase the correlation 
$\langle \Delta_{jm}^{\dagger\tilde{j}m}(i) \Delta_{jm}^{\tilde{j}m}(i^{'}) \rangle$ 
develops staggered ordering, justifying the staggered sign used below.
In Fig.~\ref{fig4}(b), we show the associated correlations summed over all distances
(with $j=1/2$ and $\tilde{j}={3/2}$),
\begin{equation}
\Delta_{m}=\frac{1}{L^{2}}\sum_{|i-i^{'}|>0} (-1)^{|i-i^{'}|}\langle \Delta_{jm}^{\dagger\tilde{j}m}(i) \Delta_{jm}^{\tilde{j}m}(i^{'})\rangle\,.
\end{equation}
Then $\Delta_{m}$ is a measure of the staggered pair-pair 
correlations associated with the EXI state.
\text

Intuitively, in the EXI phase we have hole-electron pairs involving the $(j^{eff}=3/2,\pm1/2)$
and $(1/2,\pm1/2)$ manifolds. In the absence of direct hopping between the bands, there is a conservation
of the number of electrons in each band. A nonzero expectation value for 
$\Delta_{jm}^{\tilde{j}m}(i)$ (which becomes an order parameter for this case)
amounts to a spontaneous symmetry breaking of the $U(1)$ symmetry that corresponds to the
relative phase of the bands in which the electron-hole pair forms~\cite{Kunes01}. However,
because we are using non-cubic bands with a crystal-field splitting,
this symmetry is explicitly broken in our Hamiltonian, namely in the
tight-binding term transformed to the ``$a$'' basis there is a direct hopping between
the $(3/2,\pm1/2)$ and $(1/2,\pm1/2)$ bands. Thus, 
it is somewhat surprising that the expectation values used in our work 
(like $\Delta_{m}$) still behave in practice similarly as the 
true order parameter used in~\cite{Sato01,Sato02}, 
namely it is robust in the EXI phase and very
small in other phases [see Fig.~\ref{fig4}(b)].

We also found that the staggered excitonic condensate
is always present in combination with AFM spin ordering, as deduced from the spin
structure factor $S(q)$ and the real-space spin-spin correlations
presented in Fig.~\ref{fig5}(b,d) at $\lambda/W=0.23$. We also carried out finite-size
scaling of $S(q)$ for system sizes $L$ = 8, 16 , 24, and 32
at $\lambda/W=0.046$ and $0.23$. We noticed a fast growth in the peak value at $q=\pi$ for $\lambda/W=0.23$,
indicating that spin antiferromagnetism and excitonic order are linked together. 
This aspect of stabilizing antiferromagnetism in an EXI phase due to robust Hund's coupling, 
as used by us, was discussed before in~\cite{Kaneko01}. 
Exploring the effects of varying the Hund's coupling in our model can be carried out in future work.

\begin{figure}[!th]
\hspace*{-0.0cm}
\vspace*{-3.0cm}
\begin{overpic}[width=\columnwidth]{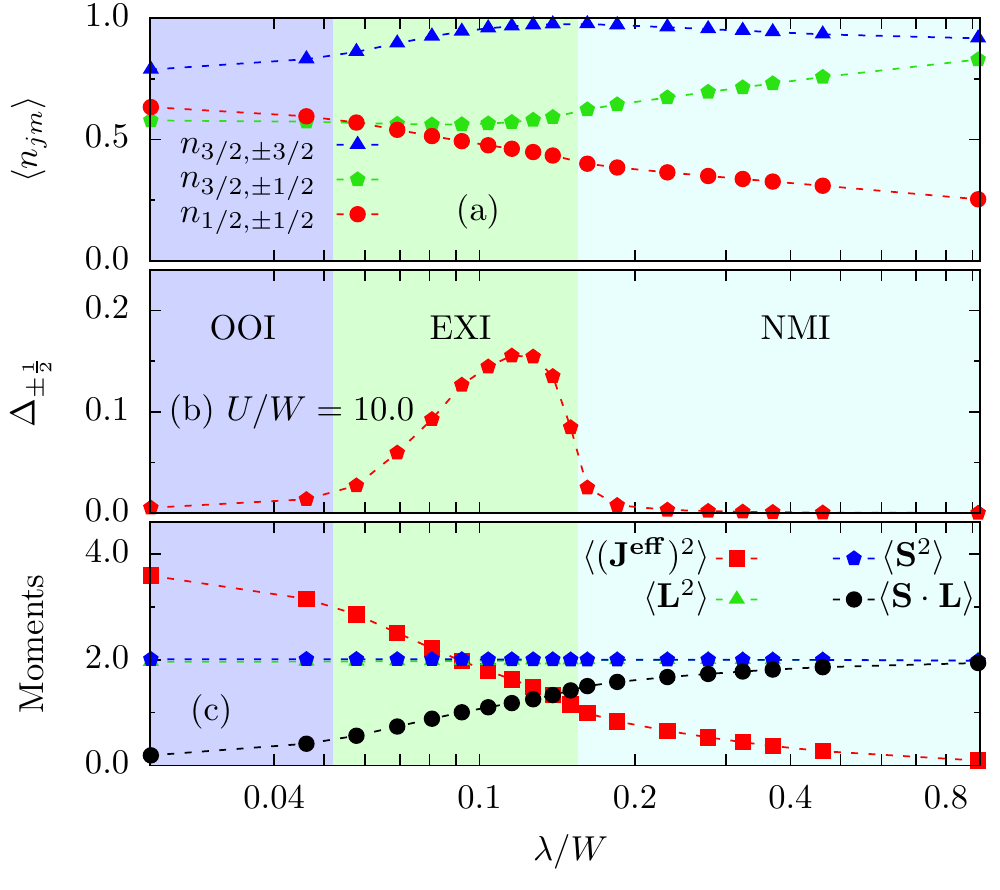} 
\end{overpic}
\begin{overpic}[width=\columnwidth]{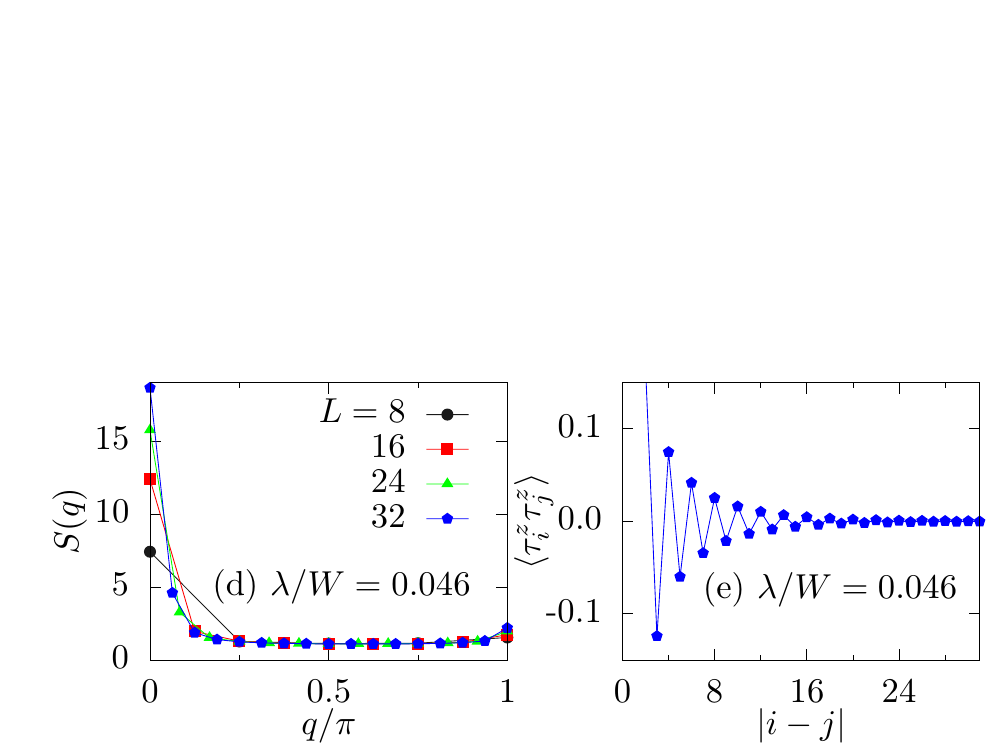} 
\end{overpic}
\caption{DMRG results obtained at $U/W=10$ (strong coupling regime). Panel (a) shows 
the occupation number in the $(j^{eff},m)$ bands, 
while panel (b) displays the excitonic order parameter dependence on $\lambda/W$. Panel (c) displays
the local moment strengths and also $\langle{\textbf{S.L}}\rangle$. Panel (d) contains 
the spin structure factor for a number of sites $L$ equal to 8~(black), 
16~(red) , 24~(green), and 32~(blue). Panel (e) shows $\langle \tau_{z}(i)\tau_{z}(j)\rangle$ for $L$=32. 
Both (d) and (e) are in the ferromagnetic and orbitally ordered phase at $\lambda/W=0.046$.}
\label{fig6}
\end{figure}

\subsection{Strong Coupling}\label{strong_coupling}

Consider now the large $U/W$ limit. In Fig.~\ref{fig6}(a-e) 
we present some results obtained at $U/W$=$10$. 
At small $\lambda/W$, we found a robust ferromagnetic (FM) spin order as shown in Fig.~\ref{fig6}(d) via the spin structure factor. 
We also noticed that this FM spin ordering is always accompanied by orbital ordering, as discussed in previous investigations 
in the absence of spin-orbit coupling~\cite{Shaozhi01}. To clarify the nature of the orbital order, 
we show $\langle\tau_{z}({{i}})\tau_{z}({{j}})\rangle$ in Fig.~\ref{fig6}(e),  
where $\tau_{z}({{i}}) = n_{{ i}yz}-n_{{ i}xz}$ is the $z$ component of the pseudospin operator in orbital space. 
This orbital ordering leads to the opening of a gap in the system rendering the state an orbital-ordered insulator (OOI), 
as discussed in~\cite{Shaozhi01}  
via determinant Quantum Monte Carlo and DMRG calculations without spin-orbit 
coupling.

\begin{figure}[!th]
\hspace*{-0.0cm}
\begin{overpic}[width=\columnwidth]{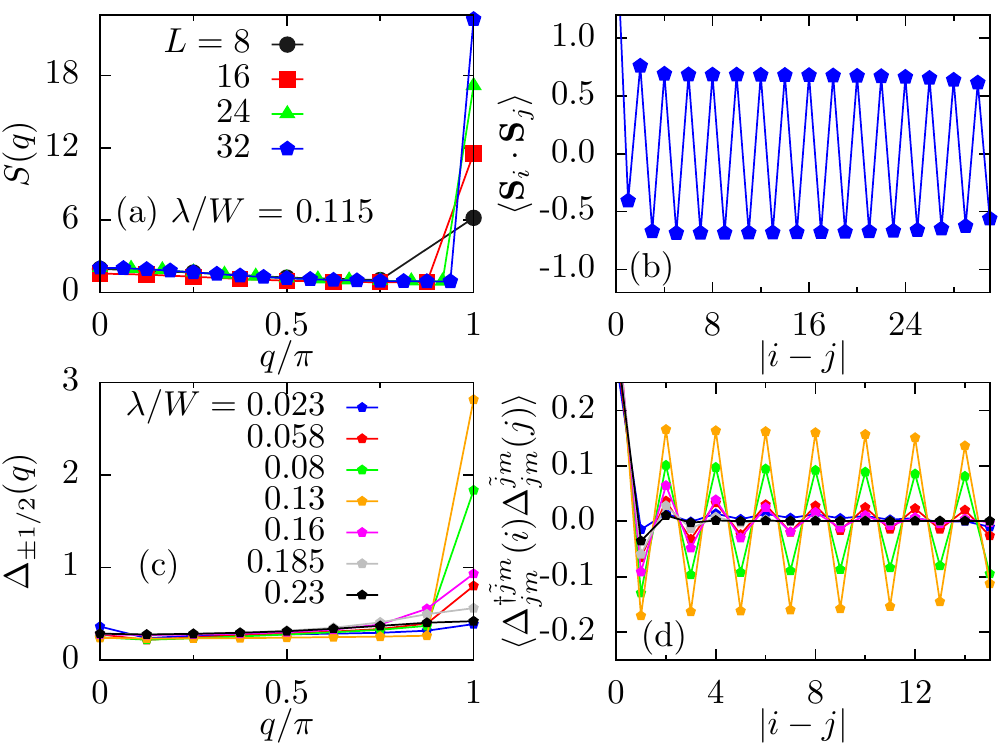} 
\end{overpic}
\caption{DMRG results obtained at $U/W$=10. Panel (a) depicts the spin structure factor in the EXI phase 
at $\lambda/W$=$0.115$, for a number of sites $L$ equal to 8~(black), 16~(red), 24~(green), and 32~(blue). 
Panel (b) shows the real-space averaged spin-spin correlations for $\lambda/W$=$0.115$. In (c) and (d), 
we present the pair-pair correlation in momentum and real space, respectively,  
for a $L=16$ system. In panel~(d) $j=1/2$, $\tilde{j}=3/2$, and $m=\pm1/2$ were used. } 
\label{fig7}
\end{figure}

By increasing $\lambda/W$, we have observed a transition from FM to the AFM
spin ordering shown in Fig.~\ref{fig7}(a,b). As in Sec.~III.B,  
this AFM ordering is accompanied by staggering 
in the exciton pair-pair correlation as shown in Fig.~\ref{fig7}(c,d). 
Similar phases were noticed in a study of the low-energy effective 
Hamiltonian for the $(t_{2g})^{4}$ sector in~\cite{Svoboda01}.  
Note that at $U/W=10$ the excitonic condensate starts at smaller $\lambda/W$ than those needed 
at intermediate value of $U/W$ [Fig.~\ref{fig6}(b)]. 
Interestingly, in the EXI phase we have noticed that $\langle{n_{3/2,\pm 3/2}}\rangle$
converges to $\approx 1$ (to be precise 0.98) 
and then reverses the trend and starts decreasing in the region identified as a non-magnetic insulator (see below). This is different from the properties of the EXI
phase observed at intermediate $U/W$ where $\langle{n_{3/2,\pm 3/2}}\rangle < 1$ 
and then slowly converged to $1$ as the system evolves
to become a band insulator increasing $\lambda/W$ further.

At $U/W$=$10$, and at any $\lambda/W$, we also noticed that $\langle {\bf S}^2 \rangle=2$ 
and $\langle {\bf L}^2 \rangle=2$ [Fig.~\ref{fig6}(c)]. These vector 
operators become parallel only for large $\lambda/W$, namely where \mbox{$\langle {\bf S}\cdot{\bf L}\rangle$=$2$} 
and $\langle ({\bf{J}}^{eff})^{2}\rangle $=$0$. 
In Fig.~\ref{fig7}(d) we show the pair-pair excitonic correlation as a function of distance, involving the operator
 $\Delta_{j,m}^{\tilde{j},m}({{i}})={a_{{{i}}\tilde{j}m}^{\dagger}}{a_{{{i}}{j}m}}^{\phantom{\dagger}}$. 
In all points studied inside the EXI phase we observed a staggering in the pair-pair correlation. 
In Fig.~\ref{fig7}(c), we show $\Delta_{m}(q) = \frac{1}{L}\sum_{i,j}e^{q(i-j)}\langle \Delta_{jm}^{\dagger\tilde{j}m}(i) \Delta_{jm}^{\tilde{j}m}(j) \rangle$ for various $\lambda$'s at strong $U$, 
where $q$ is the momentum.

\begin{figure}[!th]
\hspace*{-0.0cm}
\begin{overpic}[width=\columnwidth]{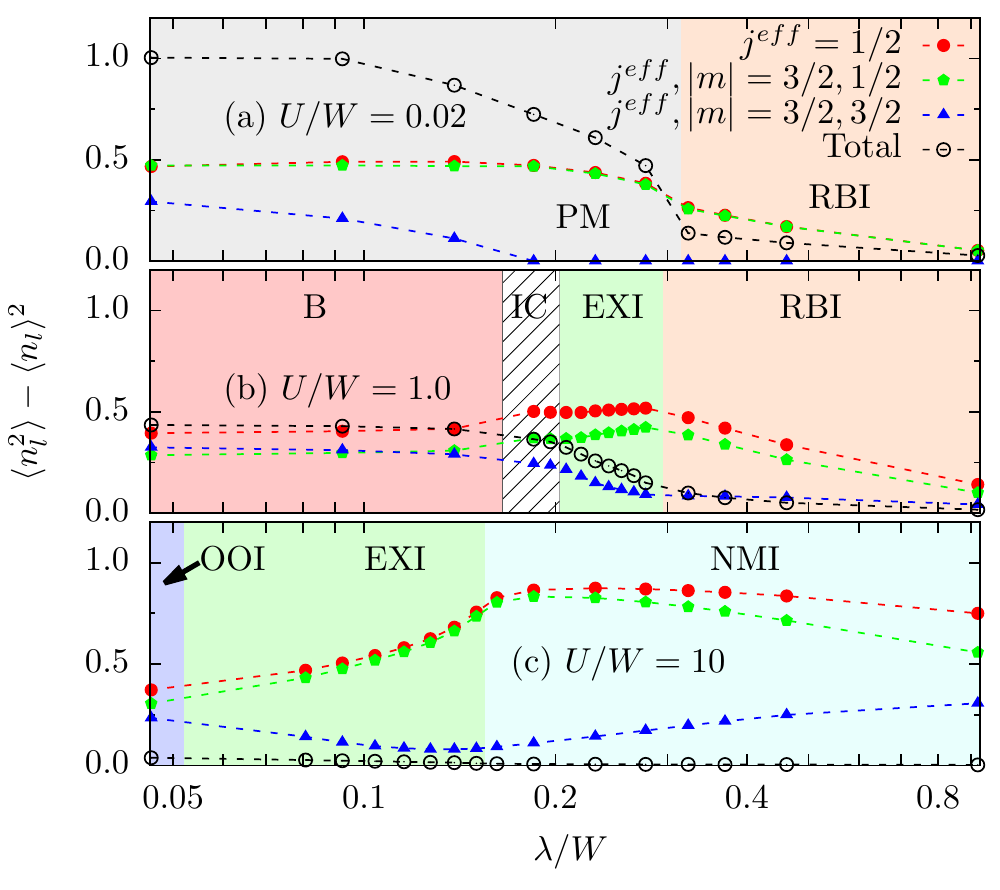}
\end{overpic}
\caption{Averaged local charge fluctuations of 
the $(j^{eff},m)$ states (as indicated in the upper panel legend) 
corresponding to (a) $U/W=0.02$ (weak coupling), (b) $U/W=1$ (intermediate coupling), and (c) $U/W=10$ (strong coupling).}
\label{fig8}
\end{figure}

In Fig.~\ref{fig8} we present $(\langle{n}^{2}_{l}\rangle - {\langle{n_{l}}\rangle}^{2})$
=$\frac{1}{L}\sum_{i}\langle{n}^{2}_{{i},l}\rangle - {\langle{n_{{i},l}}\rangle}^{2}$, 
where the index $l$ takes 
the values indicated in the legend of panel (a), namely $j^{eff}=1/2$, $(j^{eff},|m|)=(3/2,1/2)$, $(j^{eff},|m|)=(3/2,3/2)$, and $Total$ ($n_{Total}=\sum_{i,j,m}n_{ijm}$).
Interestingly, we noticed that in the EXI phase the charge fluctuations increase for $j^{eff}=1/2$ and $(j^{eff},|m|)=(3/2,1/2)$: 
these are the bands where excitons are located, and this feature is common for both intermediate and strong coupling EXI regimes. 
However, we have identified some differences within the EXI region between the intermediate and strong Hubbard coupling regions. 
In strong coupling [Fig.~\ref{fig8}(c)] we noticed that in the EXI regime the local charge fluctuations per site are nearly zero, 
suggesting that electrons are almost localized. However, at intermediate coupling [$U/W=1$, Fig.~\ref{fig8}(b)] and still within the 
EXI regime, the total charge fluctuations are nonzero. Nonzero local charge fluctuations in the EXI phase hints 
towards exciton pairs that are extended in size, namely the BCS type limit of excitonic phases.
In the other extreme, namely the Bose-Einstein condensation (BEC) limit,
the excitonic phase should have small charge fluctuations because 
the exciton pairs are considerably smaller and of atomic-scale size. 
A more detailed study of the BCS-BEC crossover 
within the excitonic phase when moving from intermediate to strong coupling $U/W$ is currently in progress.

In the strong coupling region of focus here and in the neighborhood of the EXI phase we have found a non-magnetic insulator (NMI) 
with $\langle ({\bf{J}}^{eff})^{2} \rangle \ne 0$. Let us contrast the NMI and RBI regions. To identify the NMI regime we focused 
on two aspects: {\it{(i)}} the system should have localized electrons due to strong correlations; {\it (ii)} there is no 
magnetic ordering. The first condition was checked by calculating local charge fluctuations, as shown in Fig.~\ref{fig8}, 
where we observed that the local charge fluctuations are zero throughout this region [Fig.~\ref{fig8}(c)]. This is merely 
a strong correlation effect different from the case of the small $U/W$ and large $\lambda/W$ regime (RBI) where electrons are 
primarily in extended states but still having zero local charge fluctuations because of having 
nearly full and empty bands. 
As depicted in Fig.~\ref{fig8}(a,b), in the RBI region the 
local charge fluctuations separately in each ($j^{eff},m$) 
state  as well as $Total$ are small or nearly zero. 
On the other hand, in the NMI region only $Total$ is zero but charge 
fluctuations separately for each ($j^{eff},m$) are large [shown in Fig. \ref{fig8}(c)]. This 
suggests that in the case of NMI the electrons are not locked just as the consequence 
of having a fully filled band or an empty band like in RBI, but 
as a consequence of strong correlations.

In recent work using DMFT~\cite{Sato02} 
for a cubic $(t_{2g})^4$ system, a  $\langle ({\bf{J}}^{eff})^{2} \rangle$=$0$ 
NMI state was also found in the vicinity of the excitonic 
condensate and it was identified as a Van Vleck-type Mott insulator, as discussed earlier in~\cite{Khaliullin}. Our finding 
of a NMI state with $\langle ({\bf{J}}^{eff})^{2} \rangle \ne 0$ 
near the excitonic condensate seems in contrast with those previous studies, 
but it is merely a consequence of using a non-cubic band structure. Interestingly, 
this breakdown of the $\langle ({\bf{J}}^{eff})^{2} \rangle$=$0$ 
state was also recently noticed in first-principle 
calculations~\cite{Bhowal01,Phelan01}  applied to the $(t_{2g})^{4}$ iridate Sr$_{2}$YIrO$_{6}$.  
In our results, and to the best of our accuracy, the $\langle ({\bf{J}}^{eff})^{2} \rangle \ne 0$  NMI region 
is smoothly connected to the $\langle ({\bf{J}}^{eff})^{2} \rangle$=$0$ region.

\section{Conclusions}\label{discussion}

In this publication, using an accurate computational technique we have studied the phase
diagram of an electronic model simultaneously with Hubbard, Hund, and spin-orbit couplings. The hopping
amplitudes were fixed to those used in a previous study at $\lambda =0$, since that 
effort already unveiled a variety of interesting phases such as the OSMP regime. In the
present analysis our main result is shown in Fig.~\ref{fig2}. The previously identified Block and FM-OO phases
were followed increasing $\lambda$. Eventually, over a broad range of $U/W$ an excitonic condensate
phase was identified, in qualitative agreement with previous DMFT studies. The large $\lambda$
regime is also interesting, with a variety of insulating regions.

Conceptually, our analysis provides an avenue to study quasi-one dimensional materials with robust
spin-orbit coupling. We provide a tentative partial list of materials of this class
in the introduction and throughout the text. In combination with $ab$-$initio$
techniques, needed for the hopping amplitudes, our many-body procedure can unveil properties
of these systems in reduced dimensionality with good precision to guide experiments.  We hope our
work triggers the cross-fertilization between theory and experiments needed to develop the novel
field of quasi-one dimensional iridates, or other related low-dimensional 
materials with robust spin-orbit coupling.

\section{acknowledgments}
The authors acknowledge useful conversations with Prof. G. Cao. N.K. was 
supported by the National Science Foundation Grant No. DMR-1404375. 
J.H, A.N., A.M., F.R., and E.D. were supported by the US Department 
of Energy (DOE), Office of Basic Energy Sciences (BES), Materials 
Sciences and Engineering Division.  The work of G.A. was conducted 
at the Center for Nanophase Materials Science, sponsored by the 
Scientific User Facilities Division, BES, DOE, under contract with UT-Battelle.
\appendix
\section{Theorem for conservation of $J_{z}^{eff}$}\label{appendix1}
As discussed in Sec.~II, to reduce the computational cost of our numerical calculations we target specific $J_{z}^{eff}$ sectors. In order for  $J_{z}^{eff}$ to become a good quantum number, namely to achieve $[H,J_{z}^{eff}]=0$, we need to choose carefully the parameters contained in $H_{K}$ (both the crystal-field splittings and hopping parameters) so that they satisfy the constraints discussed in this Appendix. 

Below in Eq.~(\ref{new_HK}) we show $H_{K}$ explicitly written in the $a_{jm}$ basis. This is calculated simply by using the inverse transformation of Eq.~(\ref{transformation}).
\begin{widetext}
\begin{eqnarray}\label{new_HK}
 H&=&\sum_{\langle ll^{'}\rangle}
 \renewcommand{\arraystretch}{1.5}
 \begin{array}{ccc}
 \left[\begin{array}{*6{C{6.0em}}}
    a^\dagger_{l,\frac{3}{2},\frac{3}{2}}  & 
    a^\dagger_{l,\frac{3}{2},-\frac{1}{2}} & 
    a^\dagger_{l,\frac{1}{2},-\frac{1}{2}} & 
    a^\dagger_{l,\frac{3}{2},-\frac{3}{2}} & 
    a^\dagger_{l,\frac{3}{2},\frac{1}{2}}  & 
    a^\dagger_{l,\frac{1}{2},\frac{1}{2}} \\
  \end{array}\right] &  \\
  \left[\begin{array}{cccccc}
   \frac{t_{00}+t_{11}}{2} & \frac{i t_{00}-i t_{11}+2t_{01}}{2\sqrt{3}} & \frac{-i t_{00}+i t_{11}-2t_{01}}{\sqrt{6}} & 0 & \frac{t_{12}+i t_{02}}{\sqrt{3}} & \frac{t_{12}+i t_{02}}{\sqrt{6}}  \\ 
   \frac{-i t_{00}+i t_{11}+2t_{01}}{2\sqrt{3}} & \frac{t_{00}+t_{11}+4t_{22}}{6} & \frac{-t_{00}-t_{11}+2t_{22}}{3\sqrt{2}}     & \frac{t_{12}+i t_{02}}{\sqrt{3}} & 0 & \frac{t_{02}+i t_{12}}{\sqrt{2}}  \\
   \frac{i t_{00}-i t_{11}-2t_{01}}{\sqrt{6}} & \frac{-t_{00}-t_{11}+2t_{22}}{3\sqrt{2}} & \frac{t_{00}+t_{11}+t_{22}}{3}        & \frac{t_{12}+i t_{02}}{\sqrt{6}} & \frac{-t_{02}-i t_{12}}{\sqrt{2}} & 0 \\
   0 & \frac{t_{12}-i t_{02}}{\sqrt{3}} & \frac{t_{12}-i t_{02}}{\sqrt{6}}  & \frac{t_{00}+t_{11}}{2} & \frac{i t_{00}-i t_{11}+2t_{01}}{2\sqrt{3}} & \frac{-i t_{00}+i t_{11}-2t_{01}}{\sqrt{6}} \\ 
   \frac{t_{12}-i t_{02}}{\sqrt{3}} & 0 & \frac{-t_{02}-i t_{12}}{\sqrt{2}} & \frac{-i t_{00}+i t_{11}+2t_{01}}{2\sqrt{3}} & \frac{t_{00}+t_{11}+4t_{22}}{6} & \frac{-t_{00}-t_{11}+2t_{22}}{3\sqrt{2}}     \\
   \frac{t_{12}-i t_{02}}{\sqrt{6}} & \frac{t_{02}-i t_{12}}{\sqrt{2}} & 0 & \frac{i t_{00}-i t_{11}-2t_{01}}{\sqrt{6}} & \frac{-t_{00}-t_{11}+2t_{22}}{3\sqrt{2}} & \frac{t_{00}+t_{11}+t_{22}}{3}        \\
 \end{array}\right] &
 \left[\begin{array}{cccccc}
   a_{l^{'},\frac{3}{2},\frac{3}{2}}  \\ 
   a_{l^{'},\frac{3}{2},-\frac{1}{2}} \\
   a_{l^{'},\frac{1}{2},-\frac{1}{2}} \\
   a_{l^{'},\frac{3}{2},-\frac{3}{2}} \\ 
   a_{l^{'},\frac{3}{2},\frac{1}{2}}  \\
   a_{l^{'},\frac{1}{2},\frac{1}{2}}  \\
 \end{array}\right]\end{array}
 \nonumber\\ \nonumber\\ \nonumber\\
 &+&\sum_l
 \renewcommand{\arraystretch}{1.5}
 \begin{array}{ccc}
 \left[\begin{array}{*6{C{5.15em}}}
    a^\dagger_{l,\frac{3}{2},\frac{3}{2}}  & 
    a^\dagger_{l,\frac{3}{2},-\frac{1}{2}} & 
    a^\dagger_{l,\frac{1}{2},-\frac{1}{2}} & 
    a^\dagger_{l,\frac{3}{2},-\frac{3}{2}} & 
    a^\dagger_{l,\frac{3}{2},\frac{1}{2}}  & 
    a^\dagger_{l,\frac{1}{2},\frac{1}{2}} \\
  \end{array}\right] &  \\
  \left[\begin{array}{cccccc}
   \frac{\Delta_{0}+\Delta_{1}}{2} & \frac{i\Delta_{0}-i\Delta_{1}}{2\sqrt{3}} & \frac{-i\Delta_{0}+i\Delta_{1}}{\sqrt{6}}           & 0 & 0 & 0 \\ 
   \frac{-i\Delta_{0}+i\Delta_{1}}{2\sqrt{3}} & \frac{\Delta_{0}+\Delta_{1}+4\Delta_{2}}{6} & \frac{-\Delta_{0}-\Delta_{1}+2\Delta_{2}}{3\sqrt{2}} & 0 & 0 & 0 \\
   \frac{i\Delta_{0}-i\Delta_{1}}{\sqrt{6}} & \frac{-\Delta_{0}-\Delta_{1}+2\Delta_{2}}{3\sqrt{2}} & \frac{\Delta_{0}+\Delta_{1}+\Delta_{2}}{3}    & 0 & 0 & 0 \\
   0 & 0 & 0 & \frac{\Delta_{0}+\Delta_{1}}{2} & \frac{i\Delta_{0}-i\Delta_{1}}{2\sqrt{3}} & \frac{-i\Delta_{0}+i\Delta_{1}}{\sqrt{6}}            \\ 
   0 & 0 & 0 & \frac{-i\Delta_{0}+i\Delta_{1}}{2\sqrt{3}} & \frac{\Delta_{0}+\Delta_{1}+4\Delta_{2}}{6} & \frac{-\Delta_{0}-\Delta_{1}+2\Delta_{2}}{3\sqrt{2}}  \\
   0 & 0 & 0 & \frac{i\Delta_{0}-i\Delta_{1}}{\sqrt{6}} & \frac{-\Delta_{0}-\Delta_{1}+2\Delta_{2}}{3\sqrt{2}} & \frac{\Delta_{0}+\Delta_{1}+\Delta_{2}}{3}     \\
 \end{array}\right] &
 \left[\begin{array}{cccccc}
   a_{l,\frac{3}{2},\frac{3}{2}}  \\ 
   a_{l,\frac{3}{2},-\frac{1}{2}} \\
   a_{l,\frac{1}{2},-\frac{1}{2}} \\
   a_{l,\frac{3}{2},-\frac{3}{2}} \\ 
   a_{l,\frac{3}{2},\frac{1}{2}}  \\
   a_{l,\frac{1}{2},\frac{1}{2}}  \\
 \end{array}\right]\end{array}\,.
\end{eqnarray}

\end{widetext} 
The $J_{z}^{eff}$ operator can also be written in the same basis as
\begin{equation}\label{Jzeff}
J_{z}^{eff}=\sum_{{{i}}jm}(m)n_{{i},j,m}\,.
\end{equation}
Below are the constraints on the $H_{K}$ parameters (asumming that the $t_{\gamma\gamma^{'}}$ and $\Delta_{\gamma}$ are real) which are 
obtained after imposing explicitly the condition $[H_{K},J_{z}^{eff}]=0$:
\begin{itemize}
\item {{${t_{\gamma\gamma^{'}}=0}$   $\mathbf{\forall}$ $\mathbf{\gamma \ne \gamma^{'}}$, i.e., no interorbital hopping,}}
\item ${t_{00}=t_{11}}$, namely the hopping amplitudes of the $d_{xz}$ and $d_{yz}$ orbitals must be equal,
\item ${\Delta_{0}=\Delta_{1}}$,  namely the crystal-field splittings for the $d_{xz}$ and $d_{yz}$ orbitals must be equal.
\end{itemize}
We have selected the parameters in $H_{K}$ such that the above constraints are satisfied. These constraints forbid all scattering processes of electrons under which $J_{z}^{eff}$ changes, but the hybridization between the states $(j^{eff}=3/2,m=\pm1/2)$ and $(j^{eff}=1/2,m=\pm1/2)$ 
is still allowed and our choice of parameters hybridize the above 
mentioned bands. For this reason $[({\bf{J}}^{eff})^{2},H_{K}] \ne 0$.
\vspace{0.5cm}
\section{``Good Basis'' for $H_{K} + H_{SOC}$}\label{appB}
In principle we can write the basis in which $H_{K} + H_{SOC}$ are diagonalized simultaneously. 
We name these new basis operators as $\tilde{a}_{k, \alpha, s}$; where $k$ is the momentum, 
$\alpha$ is the band index, and $s$ is the flavour of the particle, i.e., $\pm1$. Below is the relation between these new basis and $a_{k,j,m}$, where $a_{k,j,m}=(1/\sqrt{L})\sum_{l}e^{ilk}a_{l,j,m}$:
\begin{equation}
\tilde{a}_{k,0,s}^{\dagger}=a_{k,\frac{3}{2},\frac{3s}{2}}^{\dagger}\,,
\end{equation}
\begin{equation}
\tilde{a}_{k,1,s}^{\dagger}=\frac{1}{N_{2}(k)}a_{k,\frac{3}{2},\frac{s}{2}}^{\dagger} + \frac{1}{N_{1}(k)}a_{k,\frac{1}{2},\frac{s}{2}}^{\dagger}\,,
\end{equation}
\begin{widetext}
\begin{eqnarray}
\tilde{a}_{k,2,s}^{\dagger}&=&\frac{(\epsilon_{21}(k) - 9\lambda/2) + 3\sqrt{(\epsilon_{21}-\lambda/2)^{2} + 2\lambda^{2}}}{2\sqrt{2}\epsilon_{21}(k)N_{2}(k)}a_{k,\frac{3}{2},\frac{s}{2}}^{\dagger} 
+ \frac{(\epsilon_{21}(k) - 9\lambda/2) - 3\sqrt{(\epsilon_{21}-\lambda/2)^{2} + 2\lambda^{2}}}{2\sqrt{2}\epsilon_{21}(k)N_{1}(k)}a_{k,\frac{1}{2},\frac{s}{2}}^{\dagger} \,,\\
N_{\alpha}(k)&=&\frac{3((\epsilon_{21}(k)-\lambda/2)^{2} + 2\lambda^{2})^{1/4}(\sqrt{(\epsilon_{21}(k)-\lambda/2)^{2} + 2\lambda^{2}} + (-1)^{\alpha}(\epsilon_{21}(k)/3 - 3\lambda/2) )^{1/2}}{2\epsilon_{21}(k)}\,,\nonumber
\end{eqnarray}
\end{widetext}
where in the equations above, $\epsilon_{21}(k)=\epsilon_{2}(k) -\epsilon_{1}(k)$.

Using these relations, we calculated the bands 
for the non-interacting case and the $\lambda_{c}(U=0)$ 
for metal-insulator transition, as discussed in Sec.~\ref{Paramagnetic metal and Relativistic band insulator}.

\section{OSMP in the intermediate and strong $U$ coupling limit}\label{appC}
\begin{figure}[!th]
\hspace*{-0.0cm}
\begin{overpic}[trim=0 0 0 130,clip,width=\columnwidth]{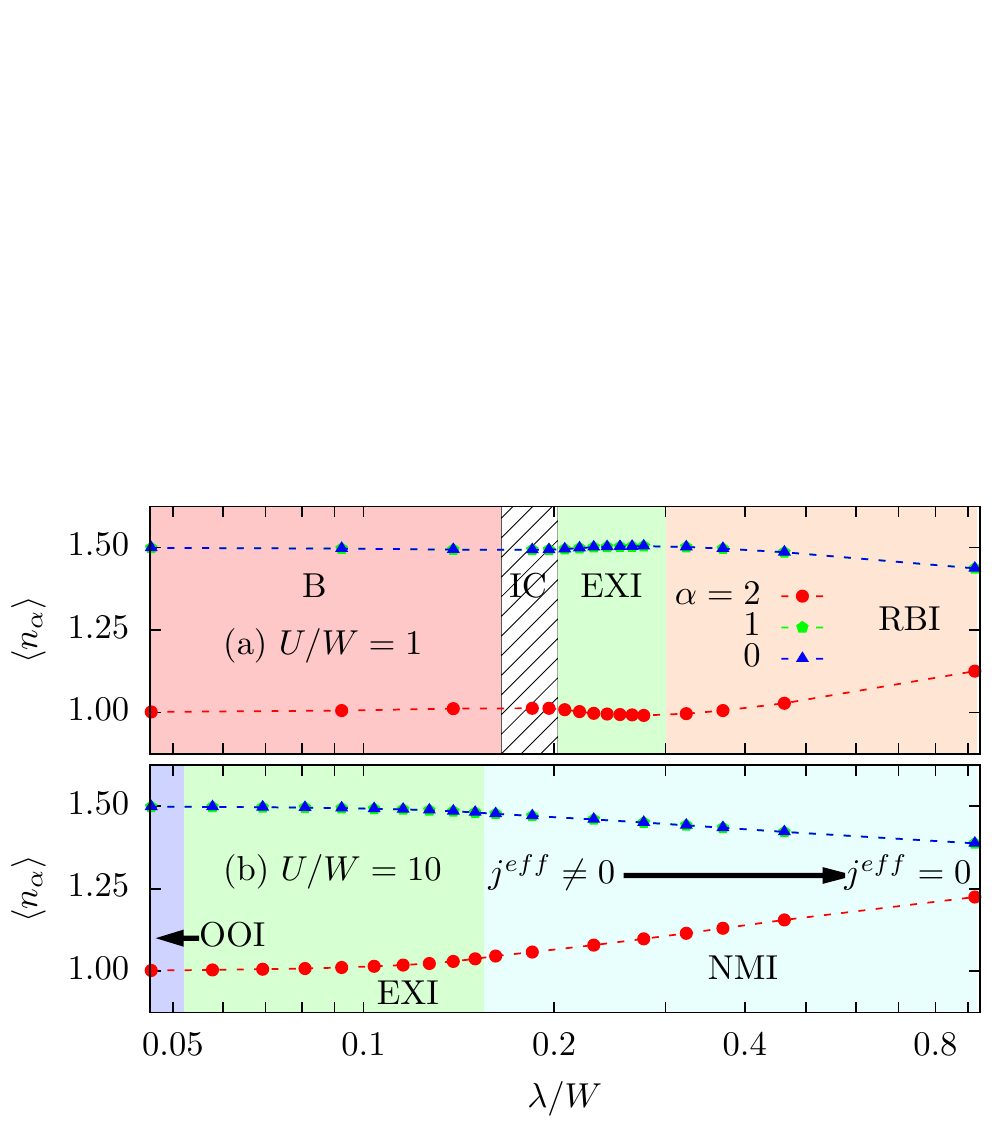}
\end{overpic}
\caption{Occupations of the $t_{2g}$ orbital states corresponding 
to (a) $U/W=1$ (intermediate coupling) and (b) $U/W=10$ (strong coupling). To a good approximation, the
excitonic condensate phase behaves similarly as the Block (OSMP) phase, 
namely with one orbital having occupation
of approximately one electron.}
\label{fig9}
\end{figure}

As discussed earlier we found the OSMP in the 
intermediate and strong coupling regions at small $\lambda$, 
by calculating occupation densities in the $t_{2g}$ basis. In the OSMP 
region the $d_{xy}$ orbital is filled with nearly one electron per site 
while $d_{xz(yz)}$ have nearly 1.5 filling.
As shown in Fig.~\ref{fig9}, we noticed that the EXI regime 
starts appearing at relatively lower values of $\lambda$ in the strong $U$ coupling region.

\end{document}